\documentclass[prl,showpacs,twocolumn]{revtex4}

\newcommand {\etal}{{\it et al.}}
\newcommand {\ACO}{$A$CrO$_2$}
\newcommand {\CrThree}{Cr$^{3+}$}
\newcommand {\CuCrO}{CuCrO$_2$}
\newcommand {\AgCrO}{AgCrO$_2$}
\newcommand {\LiCrO}{LiCrO$_2$}
\newcommand {\NaCrO}{NaCrO$_2$}

\newcommand {\CFAO}{CuFe$_{1-x}$Al$_{x}$O$_2$}
\newcommand {\Rbarm}{$R\bar{3}m$}
\newcommand {\Tn}{$T_{\mathrm{N}}$}
\newcommand {\SidotSj}{${\bf S}_i \cdot {\bf S}_j$}
\newcommand {\LiCuO}{LiCu$_2$O$_2$}
\newcommand {\TMO}{TbMnO$_3$}
\newcommand {\RFMO}{RbFe(MoO$_4$)$_2$}

\newcommand {\Spinz}{S^z_i}

\usepackage{graphicx}
\usepackage{bm}
\usepackage{pifont}
\usepackage{amsmath}

\begin{document}

\title{Spin-driven ferroelectricity and possible antiferroelectricity in triangular lattice antiferromagnets {\ACO} ($A$ = Cu, Ag, Li, or Na)}

\author{S. Seki$^1$, Y. Onose$^{1,2}$, and Y. Tokura$^{1,2}$} 
\affiliation{$^1$ Department of Applied Physics, University of Tokyo, Tokyo 113-8656, Japan \\ $^2$  Multiferroics Project, ERATO, Japan Science and Technology Agency (JST), Tokyo 113-8656, Japan}

\date{}

\begin{abstract}

Correlation between dielectric and magnetic properties was investigated on the triangular lattice antiferromagnets {\ACO} ($A=$ Cu, Ag, Li, or Na) showing 120-degree spiral spin structure with easy-axis anisotropy. For the $A=$ Cu and Ag compounds with delafossite structure, ferroelectric polarization emerges upon the spiral spin order, implying the strong coupling between the ferroelectricity and spiral spin structure. On the other hand, for the $A=$ Li and Na compounds with ordered rock salt structure, no polarization but only clear anomalies in dielectric constant can be observed upon the spiral spin order. The absence of polarization can be interpreted as the antiferroelectric state induced by the alternate stacking of {\CrThree} layer with opposite spin vector chirality. These results imply that a vast range of trigonally stacked triangular-lattice systems with 120-degree spin structure can be multiferroic, irrespective of their magnetic anisotropy.

\end{abstract}
\pacs{75.80.+q, 77.22.Ej, 75.40.Cx}
\maketitle

\begin{figure}
\begin{center}
\includegraphics*[width=5cm]{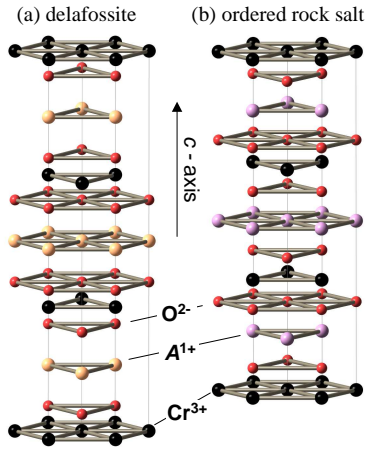}
\caption{(color online). Crystal structure of {\ACO} : (a) delafossite structure ($A=$ Cu or Ag) and (b) ordered rock salt structure ($A=$ Li or Na).}
\end{center}
\end{figure}

The correlation between magnetic and dielectric properties has long been one of the important topics in the condensed matter physics\cite{Review1, Review3}. The early attempts to realize material with both dielectric and magnetic orders (multiferroics) met difficulties, since normally these two features are mutually exclusive in their microscopic origin. Even in the rare example of multiferroics, the magnetic and dielectric phase transitions take place separately, resulting in weak coupling between both features. In a recently discovered new class of multiferroics, however, ferroelectricity arises simultaneously with the spin order\cite{Kimura}, in which magnetic (or electric) control of dielectric (magnetic) properties become possible\cite{Review2}. This group of materials is now known to commonly show magnetic frustration, which leads to complex spin order such as helimagnetic structure.

The key issue is the microscopic mechanism of coupling between ferroelectricity and magnetic order. Although the symmetry analysis of spin structure can give go/no-go rule and predict the possible direction of spontaneous polarization\cite{Harris}, thorough understanding of the microscopic origin is still lacking. So far, one of the most successful schemes to explain the behavior of ferroelectric spiral magnet is the spin-current model\cite{Katsura}, in which the electric polarization ${\bf P}_{ij}$ produced between mutually-canted magnetic moments at neighboring sites $i$ and $j$ (${\bf S}_{i}$ and ${\bf S}_{j}$) is given as
\begin{equation}
\label{KatsuraFormula}
{{\bf P}_{ij}} = A_0 \cdot {\bf e}_{ij} \times ({\bf S}_{i} \times {\bf S}_{j})
\end{equation}
Here, ${\bf e}_{ij}$ is the unit vector connecting the site $i$ and $j$, and $A_0$ a coupling constant related to the spin-orbit and spin exchange interactions. This model predicts that a helimagnet with transverse spiral components can be ferroelectric, and well explains the ferroelectric behaviors observed for $R$MnO$_3$ ($R$ = Tb and Dy) \cite{Kimura, Kimura2, Goto}, Ni$_3$V$_2$O$_8$\cite{Ni3V2O8}, CoCr$_{2}$O$_4$\cite{CoCr2O4}, MnWO$_4$\cite{MnWO4}, LiCu$_2$O$_2$\cite{LiCu2O2, seki_neutron}, LiCuVO$_4$\cite{LiCuVO4, Loidl}, and so on.

In contrast, the explanation of magnetoelectric coupling is not so straightforward for triangular lattice antiferromagnet, the most typical example of geometrically frustrated spin system. With classical Heisenberg spins, this system generally favors 120$^{\circ}$ spiral spin structure at the ground state. Depending on the sign of anisotropy term $\mathcal{H}^{'}=D \sum (\Spinz)^2$, the spin spiral is confined in the plane parallel ($D>0$ : easy-plane type) or perpendicular ($D<0$ : easy-axis type) to the triangular lattice plane\cite{Triangular}. Although in neither case can the spin-current model predict ferroelectricity, the appearance of polarization in 120$^\circ$ magnetic phase has recently been reported for {\RFMO} with easy-plane anisotropy\cite{RFMO}. This behavior can be justified by the symmetry consideration, yet leaving its microscopic origin still unclear.

The target of this paper, {\ACO} ($A$ = Cu, Ag, Li, or Na) is another example of triangular lattice antiferromagnet. {\CuCrO} and {\AgCrO} crystallize into the delafossite structure (Fig. 1 (a)). Each element forms the triangular lattice and stacks along the $c$-axis in the sequence Cr$^{3+}$-O$^{2-}$-$A^{+}$-O$^{2-}$-Cr$^{3+}$. {\LiCrO} and {\NaCrO} crystallize into the ordered rock salt structure with the similar triangular lattice (Fig. 1(b)). Both belong to the space group {\Rbarm}, and only a difference is the stacking pattern of O$^{2-}$-$A^{+}$-O$^{2-}$ layers. While the delafossite structure has the straight stacking, the ordered rock salt structure has the zigzag one. In both cases the rhombohedral (ABCABC...) stacking is realized among {\CrThree} layers, although the distance between them is much shorter in the latter case\cite{ACrO2_exchange}. The magnetic properties are dominated by {\CrThree} ion with $S=3/2$ spin, which is surrounded by octahedron of O$^{2-}$. Because of the geometrical frustration of intra-plane antiferromagnetic exchange interaction, the 120$^{\circ}$ spin structure is realized at the ground state. Based on several neutron studies as mentioned later, these systems are generally recognized to have easy-axis anisotropy along the $c$-axis. 

In this paper, we report the discovery of the spin-driven ferroelectricity and also possible antiferroelectricity by 120$^\circ$ spin structure with easy-axis anisotropy. Combined with the case for {\RFMO}\cite{RFMO}, we can predict that a vast range of trigonally stacked triangular-lattice systems with the 120$^\circ$ spin structure can be multiferroic, irrespective of their magnetic anisotropy.

Powder specimen of {\CuCrO}, {\AgCrO}, {\LiCrO}, and {\NaCrO} were prepared by solid state reaction from stoichiometric mixture of CuO, Ag, Li$_2$CO$_3$, Na$_2$CO$_3$ and Cr$_2$O$_3$. They were heated at 1000 $^{\circ}$C for 24 hours in air, at 900 $^{\circ}$C for 48 hours in O$_2$, at 1200 $^{\circ}$C for 24 hours in air, and at 1100 $^{\circ}$C for 30 hours in Ar, respectively. Powder x-ray diffraction measurements showed no detectable impurity, except slight Ag phase in {\AgCrO} specimen and slight Cr$_2$O$_3$ phase in {\NaCrO} specimen. They were pressed into rod, sintered with additional heating, and cut into thin plate. The typical sample size is 4.5mm $\times$ 4.5mm $\times$ 0.7mm. As the electrodes, silver paste was put on the widest faces. Dielectric constant was measured at 100kHz using an $LCR$ meter. To deduce the electric polarization, we measured the pyroelectric current with a constant rate of temperature sweep (2K/min$\sim$20K/min) and integrated it with time. To obtain a single ferroelectric domain, the poling electric field was applied in the cooling process and removed just before the measurements of pyroelectric current. Heat capacity was measured by the thermal relaxation method. Magnetization was measured with a SQUID magnetometer.

\begin{figure}
\begin{center}
\includegraphics*[width=7cm]{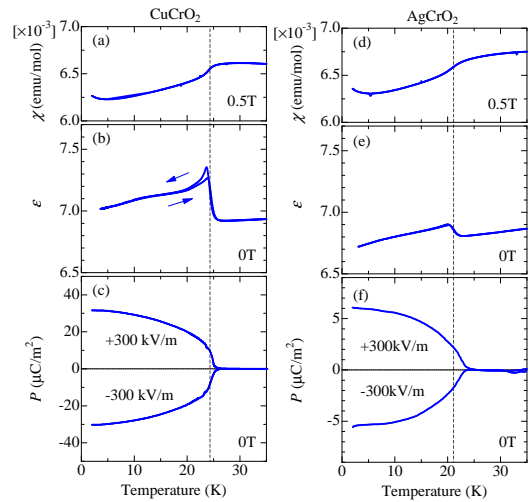}
\caption{(color online). Temperature dependence of magnetic susceptibility $\chi $, dielectric constant $\varepsilon $, and electric polarization $P$ for (a)-(c) {\CuCrO} and (d)-(f) {\AgCrO} with delafossite structure. In (c) and (f), the magnitude and sign of poling electric field are also indicated. }
\end{center}
\end{figure}

Figures 2 (a)-(c) show the temperature dependence of magnetic susceptibility, dielectric constant, and electric polarization for {\CuCrO}. The susceptibility shows a clear kink at {\Tn} $\sim$ 24K; {\Tn} is in accord with the previous report\cite{CuCrO2_mag}. A former powder neutron study suggests the 120$^\circ$ spin structure with spiral plane including $c$-axis below {\Tn}\cite{CuCrO2_neutron}. At {\Tn}, dielectric constant also shows a sharp anomaly, and spontaneous electric polarization begins to develop. With opposite poling electric field, the polarization direction can be reversed. These indicate the ferroelectric nature of the magnetic ground state, and imply the coupling between the ferroelectricity and spiral magnetic order.

Figure 3(a) indicates the symmetry elements in the {\ACO} system with space group {\Rbarm}; reflection mirror ($m$), two-fold rotation axis(2), inversion center, and three-fold rotation axis along the $c$-axis. Because of the ambiguity of spin structure, hereafter we examine two types of $120^{\circ}$ magnetic order with spin spiral either in the $(110)$ plane(Fig. 3(b)) or in the $(1\bar{1}0)$ plane(Fig. 3(c)). The former case can be considered as the proper screw magnetic structure, whose spins rotate in the plane perpendicular to the modulation vector. Recently, some specific speculation was given by Arima\cite{Arima} for this situation with the delafossite crystal structure and proper screw spin structure. With the $(110)$ spin spiral plane, only a $2'$ symmetry element, two-fold rotation axis along the [110] direction with time reversal operation, remains unbroken. Since electric polarization vector ${\bf P}$ must be invariant under the symmetry operation, only ${\bf P}$ perpendicular to the spin spiral plane (along the [110] direction) is allowed. The problem to be solved next is its microscopic origin. Because any 120$^{\circ}$ spin structure gives the same {\SidotSj} for all bonds in the regular triangular lattice, conventional magnetostriction cannot cause the net polarization with centrosymmetric crystal structure. Another candidate for the microscopic origin of ${\bf P}$ is the spin-current model or inverse Dzyaloshinskii-Moriya mechanism represented by Eq. (\ref{KatsuraFormula}). However, this also fails to explain the emergence of ferroelectricity for the regular triangular lattice. Recently, Jia {\etal} pointed out that the spin-orbit interaction brings about some modification on $d$-$p$ hybridization between ligand ion and 3$d$ magnetic ion, which can cause the polarization along the bond direction\cite{Jia1, Jia2}. Although this term oscillates in the crystal and usually cannot cause macroscopic polarization, some components along the modulation vector are proven not to be canceled out in the delafossite system with proper screw spin structure\cite{Arima}. In {\CFAO} with the same crystal structure and the proper screw spin configuration (with incommensurate wave number $q\sim0.22$), the emergence of the polarization along the modulation vector is confirmed\cite{CuFeO2,seki,Nakajima}. The similar situation is anticipated to occur in {\CuCrO}. In the case of $(1\bar{1}0)$ spiral plane(Fig. 3 (c)), on the other hand, only a reflection mirror can survive or disappear depending on the spin direction. Therefore, from the symmetry, polarization can be allowed in any direction. The spin current model predicts the polarization $(1-\alpha) P_0$ along the $c$-axis, where $\alpha$ represents the difference of coupling constant $A_0$ in Eq. (1) between chains along [110] and [100] (or [010]). Given the isotropic coupling constant ($\alpha=1$), the polarization should vanish, and hence other microscopic origin would be required. Note that similar argument as above can be constructed for other centrosymmetric trigonal systems.

\begin{figure}
\begin{center}
\includegraphics*[width=7cm]{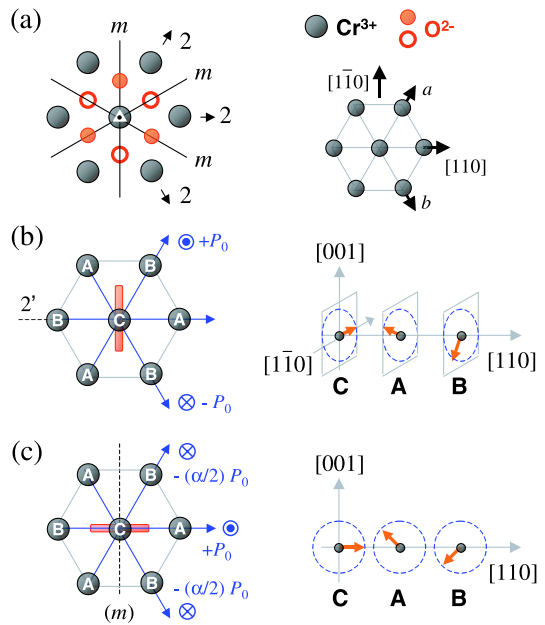}
\caption{(color online). (a) Symmetry elements in the {\ACO} system with space group {\Rbarm} : twofold rotation axis (2), reflection mirror ($m$), and threefold rotation axis with inversion center (triangle with small circle). O$^{2-}$ site above (below) the Cr$^{3+}$ layer is indicated as closed (open) circle. (b)-(c) Symmetry elements compatible to $120^{\circ}$ spin structure with (b) (110) spiral plane or (c) $(1{\bar{1}}0)$ spiral plane. The thick bars (left panel) indicate the spin spiral plane. Electric polarization expected from the spin-current model along each chain is also indicated, such as $\pm P_0$ and $-(\alpha/2)P_0$ (see text).}
\end{center}
\end{figure}

It is interesting to see a generic feature of the dielectric response in other triangular-lattice Cr-oxides. Among them, the isostructural material {\AgCrO} also shows the similar coupling between ferroelectricity and magnetic order. Figures 2 (d)-(f) indicate the temperature profiles of the same physical quantities for {\AgCrO}. The kink in magnetic susceptibility is observed at slightly lower temperature, {\Tn} $\sim $21K. Again, anomaly in dielectric constant and appearance of ferroelectric polarization ${\bf P}$ are observed at {\Tn}, although the ${\bf P}$ value is reduced as compared with {\CuCrO}. A former powder neutron study has proposed a slightly modulated 120$^\circ$ spin structure for the magnetic ground state below {\Tn}\cite{AgCrO2_neutron}. Mekata {\etal} explained this modulation by the competition between the intra-plane and inter-plane exchange interactions, and reported the shorter correlation length\cite{AgCrO2_neutron} and larger spin fluctuation\cite{AgCrO2_SR} than in {\CuCrO}. Although the detail of magnetic structure, such as the direction of spin spiral plane, has not been determined yet, the smaller spontaneous polarization value in {\AgCrO} ($\sim 1/5$ of that for {\CuCrO}) is consistent with these features. Since dielectric constant $\varepsilon $ reflects the fluctuation of polarization $\Delta P$ in the form of $\varepsilon - \varepsilon_\infty \propto  \langle |\Delta P|^2\rangle /k_\mathrm{B} T$, the weaker anomaly in $\varepsilon$ must come from the smaller polarization. 

\begin{figure}
\begin{center}
\includegraphics*[width=7cm]{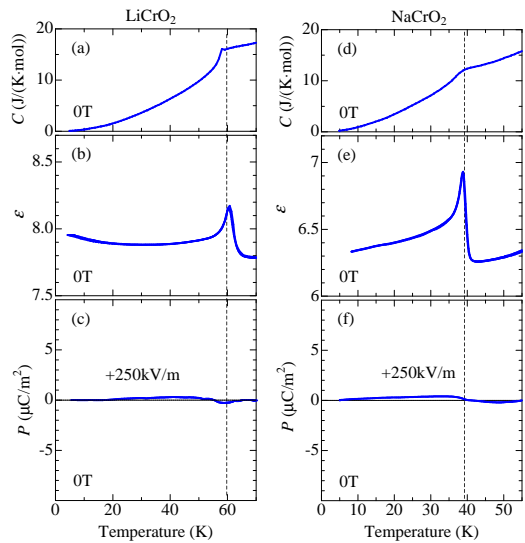}
\caption{(color online). Temperature profiles of specific heat capacity $C$, dielectric constant $\varepsilon $, and electric polarization $P$ for (a)-(c) {\LiCrO} and (d)-(f) {\NaCrO} with ordered rock salt structure. Note that the ordinate scale in (c) and (f) is the same as in Fig. 2(f).}
\end{center}
\end{figure}

In addition to the above delafossite crystals, we have also investigated {\LiCrO} and {\NaCrO} composed of the similar CrO$_2$ sheet but with ordered rock salt structure (Fig. 1(b)). The magnetic structure of {\LiCrO} has been investigated by the polarized neutron study on a single crystal\cite{LiCrO2_neutron}, and below {\Tn} $\sim$ 60K\cite{LiCrO2_mag} the proper screw type 120$^\circ$ spin structure (Fig. 3 (b)) was reported to give the best fit. For {\NaCrO}, only a powder neutron study was performed\cite{NaCrO2_neutron} and {\Tn}$\sim$ 40K has been reported\cite{NaCrO2_mag}. Figures 4 (a) - (f) indicate the temperature profiles of heat capacity, dielectric constant, and electric polarization for {\LiCrO} and {\NaCrO}. Although the anomaly in magnetic susceptibility is not clear\cite{NaCrO2_mag}, the heat capacity manifests magnetic phase transitions, as seen in Figs. 4(a) and (b), in accord with the former neutron studies. At {\Tn}, dielectric constant shows a strong cusp like anomaly as in the two compounds with delafossite structure. This suggests the large fluctuation of electric dipole around {\Tn}, and confirms the correlation between dielectric and magnetic natures also in this system. However, unlike the case for delafossites of {\CuCrO} and {\AgCrO}, the macroscopic polarization can hardly be observed for {\LiCrO} or {\NaCrO}. One of the possible interpretations for the absence of {\bf P} but the presence of sharp $\varepsilon$-peak is the antiferroelectric order of electric dipoles. For {\LiCrO}, on the basis of the two magnetic modulation vectors ${\bf q}_1=(1/3, 1/3, 0)$ and ${\bf q}_2=(-2/3, 1/3, 1/2)$, alternate stacking  of {\CrThree} layer with opposite vector spin chirality was suggested\cite{LiCrO2_neutron}. Since recent polarized neutron studies on several multiferroics (such as {\TMO}\cite{Yamasaki}, {\CFAO}\cite{Nakajima}, and {\LiCuO}\cite{seki_neutron}) confirm the coupling between the spin helicity and the sign of polarization, it is natural to consider such an antiferro-chiral order leads to the antiferroelectric state. For {\CuCrO} and {\AgCrO}, by contrast, the ${\bf q}_2$ peaks, which characterize the alternate stacking of opposite chirality layers, have not been observed in neutron profiles\cite{CuCrO2_neutron, AgCrO2_neutron} in accord with the emergence of ferroelectricity in these compounds. The absence of polarization in {\LiCrO} can conversely suggest that in {\ACO} system the spin helicity determines the direction of polarization. At this stage, the origin of interaction that stabilizes such antiferro-chiral spin order is an open question, but may possibly be ascribed to the inter-layer magnetic and/or electrostatic interaction. The different stacking pattern of O-$A$-O layers and shorter distance between {\CrThree} layers, which is anticipated to cause stronger inter-plane interaction and higher {\Tn}\cite{ACrO2_exchange}, may be related to the antiferroic order of spin chirality. Further in general, the antiparallel arrangement of {\bf P} will be favored between the in-plane ferroelectric sheets, which may in turn make the stacking of spin vector chirality antiferroic.

In summary, we investigated the correlation between dielectric and magnetic properties of triangular lattice antiferromagnet {\ACO} showing 120$^\circ$ spin structure with easy-axis anisotropy. For the $A=$ Cu and Ag compounds with delafossite structure, appearance of electric polarization was observed concurrently with the magnetic order, implying the strong magnetoelectric coupling in this system. For the $A=$ Li and Na compounds with ordered rock salt structure, by contrast, no polarization but only anomalies in dielectric constant were observed at {\Tn}. Considering the results of the former neutron study, this can be interpreted as the antiferroelectric state due to the alternate stacking of magnetic layers with opposite spin vector chirality. Combined with the recent results for {\RFMO} with easy-plane anisotropy\cite{RFMO}, a vast range of trigonally stacked triangular-lattice systems with 120$^\circ$ spin structure can be multiferroic, irrespective of their magnetic anisotropy.

The authors thank T. Arima, Y. Yamasaki, H. Katsura, S. Tanaka, R. Kumai, S. Ishiwata, and N. Nagaosa for enlightening discussions. This work was partly supported by Grants-In-Aid for Scientific Research (Grant No. 16076205, 17340104) from the MEXT of Japan.

\end{document}